\documentclass[amsfonts,12pt, a4wide,righttag, ctagsplt]{amsart}
\usepackage{enumerate}

\numberwithin{equation}{section} \overfullrule=0pt
\newtheorem{thm}{Theorem}[section]

\newtheorem{lem}[thm]{Lemma}

\newcommand{\eq}[1]{(\ref{#1})}
\newcommand{\lemm}[1]{Lemma \ref{#1}}
\newcommand{\theo}[1]{Theorem \ref{#1}}

\newcommand{\sect}[1]{Section \ref{#1}}

\newcommand{\UqX}{U^q_X}

\newcommand{\barX}{{\bar X}}
\newcommand{\uX}{{\underline X}}

\newcommand{\UqbX}{U^q_{\barX}}
\newcommand{\UquX}{U^q_{\uX}}

\newcommand{\bfo}{{\bf 1}}

\newcommand{\bbe}{\begin{equation}}

\newcommand{\INT}{\int_{-\infty}^{+\infty}}

\newcommand{\R}{{\mathbb R}}

\newcommand{\bQ}{{\mathbb Q}}

\newcommand{\cD}{{\mathcal D}}
\newcommand{\cF}{{\mathcal F}}

\newcommand{\cK}{{\mathcal K}}
\newcommand{\cM}{{\mathcal M}}

\newcommand{\eps}{\epsilon}
\newcommand{\de}{\delta}

\newcommand{\be}{\beta}
\newcommand{\bep}{\beta^+}
\newcommand{\bem}{\beta^-}

\newcommand{\bepo}{\be^+_1}
\newcommand{\bept}{\be^+_2}
\newcommand{\bemo}{\be^-_1}
\newcommand{\bemt}{\be^-_2}

\newcommand{\ap}{a^+}
\newcommand{\am}{a^-}
\newcommand{\apo}{a^+_1}
\newcommand{\amo}{a^-_1}
\newcommand{\apt}{a^+_2}
\newcommand{\amt}{a^-_2}

\newcommand{\cp}{c^+}
\newcommand{\cm}{c^-}

\newcommand{\ee}{\end{equation}}

\newcommand{\lp}{\lambda^+}
\newcommand{\lm}{\lambda^-}

\newcommand{\kapq}{\kappa^+_q}
\newcommand{\kamq}{\kappa^-_q}

\newcommand{\Om}{\Omega}

\newcommand{\om}{\omega}

\newcommand{\ka}{\kappa}
\newcommand{\sg}{\sigma}
\newcommand{\sgm}{\sigma_-}
\newcommand{\sgp}{\sigma_+}

\begin{document}
\title[Universal bad news principle]
{Universal bad news principle and pricing of options on
dividend-paying assets}

\author[S.~Boyarchenko]{Svetlana Boyarchenko$^{\ast}$}
\author[S.~Levendorski\v{i}]{Sergei Levendorski\v{i}$^\dagger$}
%\keywords{Non-Gaussian processes, perpetual Bermudan options}
%\subjclass{Primary-90A09, 60G40; \\
%secondary-- 49K05}
\maketitle

\centerline{\small $^*$ Department of Economics, The University of
Texas at Austin,} \centerline{\small 1 University Station C3100,
Austin, TX 78712, U.S.A.} \centerline{\small e-mail:
sboyarch@eco.utexas.edu}

\centerline{\small $^\dagger$ Department of Economics, The
University of Texas at Austin,} \centerline{\small 1 University
Station C3100, Austin, TX 78712, U.S.A.} \centerline{\small
e-mail: leven@eco.utexas.edu}

\begin{abstract}
We solve the pricing problem for perpetual American puts and calls
on dividend-paying assets. The dependence of a dividend process on
the underlying stochastic factor is fairly general: any
non-decreasing function is admissible. The stochastic factor
follows a L\'evy process. This specification allows us to consider
assets that pay no dividends at all when the level of the
underlying factor (say, the assets of the firm) is too low, and
assets that pay dividends at a fixed rate when the underlying
stochastic process remains in some range. Certain dividend
processes exhibiting mean-reverting features can be modelled  as
appropriate increasing functions of L\'evy processes. The pay-offs
of both the American put and call options can be represented as
the expected present value (EPV) of a certain stream of dividends:
$g(X_t)=\de(X_t)-qK$ and $g(X_t)=qK-\de(X_t)$, respectively, and
we show that the option must be exercised the first time the EPV
of the stream $g(\uX_t)$, where $\uX_t=\inf_{0\le s\le t}X_s$ is
the infimum process starting from the current level $X_0$, becomes
positive. Thus, the exercise threshold depends only on the record
setting bad news.  The results can  be applied to the theory of
real options as well; as one of possible applications, we consider
the problem of incremental capital expansion.

\end{abstract}

\section{Introduction}
The objective of this paper is to provide a general framework for
 pricing and optimal exercise strategies for American options on dividend-paying
assets, for fairly general dependence of dividend rate,
$\de(X_t)$, on
 the underlying stochastic factor. The stochastic factor, $X_t$, follows a  L\'evy process.
The standard approach uses the price of an asset as the primitive;
we show that the use of the dividend process as a primitive has
certain advantages. In the case of an asset which evolves as a
geometric Brownian motion or L\'evy process, and pays dividends as
a constant proportion of the asset's value, both specifications
can easily be  transformed one into another. The use of the
dividend process as a primitive allows us to consider assets that
pay no dividends at all when the level of the underlying factor
(say, the value of the firm) is too low, and assets that pay
dividends at a fixed rate when the underlying stochastic process
remains in some range. Certain dividend processes exhibiting
mean-reverting features can be modelled  as appropriate increasing
functions of L\'evy processes. In this paper, we consider
perpetual American put and call options; by using the variant of
Carr's randomization procedure developed in Levendorski\v{i}
(2004), which is, essentially, a sequence of embedded perpetual
options, it is possible to apply the method of the paper to the
case of American options with finite time horizon.

Given a candidate for the optimal exercise threshold, we calculate the option value, and the form of the solution
suggests the following description of the optimal exercise strategy.
 If the payoff stream is a decreasing
function of the underlying stochastic factor, then it is optimal to exercise a put-like option the first time the
EPV of the stream of payoffs calculated for the supremum process instead of the original stochastic process
becomes non-positive. Similarly, if the payoff stream is an increasing function of the underlying stochastic
factor, then it is optimal to exercise a call-like option the first time the EPV of the stream of payoffs
calculated for the infimum process instead of the original stochastic process becomes non-negative. This allows us
to formulate a general optimal exercise rule:  it is optimal to exercise the right for  (respectively, to give up)
the stream of stochastic payoffs, $g_t$ when the EPV of
 the stream $\underline{g}_t=\inf_{0\le s\le t}g_t$,  becomes non-negative (respectively,
 non-positive). We call the above statement a {\em universal record setting bad news principle}. This principle
 naturally generalizes and extends Bernanke's (1983) {\em bad news principle} and record setting news principles
spelled out in Boyarchenko (2004). In the latter paper, the principles were stated and proved for the streams of
the form $e^{X_t}-K$ and $K-e^{X_t}$, where $X_t$ is a L\'evy process. Here the result is proved for any monotone
function, $g_t=g(X_t)$. The method of the paper works for some non-monotone
payoff streams as well.

If the underlying process is a diffusion with exponentially
distributed jumps, calculation of the optimal exercise price and
rational option price reduce to calculation of simple integrals,
and solution of one equation. If, in addition, the dividend
process is a piece-wise constant function, or more generally,
piece-wise exponential polynomial, then all the integrals can be
calculated explicitly, and the optimal exercise price can be found
as a unique solution of an algebraic equation with a monotone
function.

 Now we describe the findings of the paper in more details.
Let $\de(X_t)$ be the dividend process on the asset. The riskless
rate $q>0$ is fixed.  Assume that the underlying stochastic factor
$\{X_t\}$ is a L\'evy process under a risk-neutral measure chosen
by the market, denoted $\bQ$, and let $(\Om, \cF, \bQ)$ be the
corresponding probability space (for general definitions of the
theory of L\'evy processes, see, e.g., Bertoin (1996) and Sato
(1999)). Then the EPV of the stream $g(X_t)$, at the spot level
$X_0=x$, is given by
\begin{equation}\label{epvX}
(\UqX \de)(x):=E^x\left[\int_0^\infty e^{-qt}\de(X_t)dt\right].
\end{equation}
For the perpetual American call on the asset, the payoff function
is
\[
G(x)=(\UqX \de)(x)-K=\UqX (\de(\cdot)-qK)(x), \] where $K$ is the
strike price, and for the perpetual  American put, the payoff
function is
\[ G(x)=K-(\UqX
\de)(x)=\UqX (qK-\de(\cdot))(x). \] (The standard specification is
$G(X_t)=e^{X_t}-K$ and $G(X_t)=K-e^{X_t}$, respectively, where
$X_t$ is the log-price of the stock.)
  The
rational price of the option with the payoff $G(X_t)$ is given by
\bbe\label{opt} V(x)=\sup E^x[e^{-q\tau}G(X_\tau)],
\end{equation}
where $E^x$ denotes the expectation under $\bQ$, and the supremum
is taken over a set $\cM$ of all stopping times $\tau=\tau(\om)$
satisfying $0\le \tau(\om)\le +\infty$, $\om\in\Omega$; if
$\tau(\om)=+\infty$, then $G(\tau(\om))=0$ by definition (see,
e.g., Shiryaev (1999), XVIII, 2). Notice that we use $G(X_t)$
rather than $\max\{G(X_t), 0\}$, which is admissible because the
option is not exercised unless $G(X_t)$ is positive (the
equivalence of these two specifications was used in Darling et al.
(1972)). In the paper,  the optimal stopping time, $\tau$, turns
out to be the hitting time of a semi-finite interval of the form
$(-\infty, h]$ (put-like options) or $[h, +\infty)$ (call-like
options). We denote these hitting times by $\tau^-_h$ and
$\tau^+_h$, respectively. The class of the hitting times of
semi-finite intervals is denoted by $\cM_0$.

Perpetual American options were considered by many authors, both
in discrete and continuous time models. Mc Kean (1965) calculated
the exercise boundary and price for perpetual call option in the
continuous time Gaussian model, Darling et al (1972) solved the
corresponding problem in the discrete time model, for arbitrary
random walk, and Merton (1973) solved the problem for the put in
the continuous time Gaussian model. Starting from the middle of
1990-th, a series of results for L\'evy processes of varying
degree of generality were obtained by various authors,  using
different methods (see the bibliography in Boyarchenko and
Levendorski\v{i} (2002a, b) and Mordecki (2002), and a more
detailed discussion in \sect{put}.

In the current paper, we present the solution to the optimal
stopping problem for wide class of L\'evy processes satisfying the
(ACP)--property (absolute continuity of potential kernels: see,
e.g., Sato (1999)), and fairly general payoff functions. We
formulate our results in terms of a stream of payoffs, $g$
(dividends), whose expected present value (EPV) is equal to the
given payoff, $G$ (spot-price of the stock), and {\em expected
present value operators}, $\UqbX$ and $\UquX$, of the supremum and
infimum processes
%(these operators are defined by essentially the
%same formulas as the resolvent of a Markov process;
 (these operators are defined by formulas similar to \eq{epvX}; for details,
see \sect{main}). In the case of the put on a stock which pays no
dividends and similar put-like options, we formulate the results
separately, in terms of the payoff function itself. The reason is
that the price of a stock which pays no dividends cannot be
determined as the EPV of any stream. For the case of call-like
options (the case of an increasing $G$),
 we prove the optimality in the class $\cM$, under the
weak conditions that the EPV of the stream $g$ under the infimum
process, $\UquX g$, changes sign from ``-" to ``+" only once, and
the stream $g$ is a non-decreasing function.  The last condition
is not necessary; in fact, the proof in the paper works in some
situations when the stream is not monotone. Similar results are
proved for put-like options with decreasing payoff functions $G$.
This time, the optimal exercise price is the zero of the EPV of
the stream $g$ under the supremum process, $\UqbX g$, which is
assumed to change sign from ``+" to ``-" only once.

%The key element of our approach is the representation of the option value, which is natural for the form of  the
%Wiener-Hopf factorization method used in analysis, and the scheme of the proof is essentially the same as in
%Boyarchenko and Levendorski\v{i} (2000, 2002a,b), where mainly analytical tools were used. However, the proof
%simplifies considerably and requires only the (ACP)--property thanks to the probabilistic interpretation of
%factors in the operator form of the Wiener-Hopf factorization method as the resolvent operators of the supremum
%and infimum process. After the basic properties of the resolvent operators are established, the proof of the
%optimality of the solution is less than 2 pages long.

The rest of the paper is organized as follows. In \sect{levy}, we
recall the basic definitions of the theory of L\'evy processes,
introduce the EPV operators of the supremum and infimum processes,
and calculate their action for the case of diffusions with
exponentially distributed jumps. In \sect{main} formulate the main
results, and explicitly calculate the optimal exercise thresholds
and option prices for diffusions with exponentially distributed
jumps. The proofs of the main results are in
\sect{WH}--\sect{call}. In \sect{capexp}, we apply the method of
the paper to the problem of capital expansion.

%In \sect{WH}, we recall the Wiener-Hopf factorization, introduce
%the resolvent operators of the supremum and infimum processes, and
%express the stochastic expressions of the form
%$E^x\left[e^{-q\tau}g(X_\tau)\right]$, where $\tau$ is the hitting
%time of a semi-finite interval, in terms of the resolvents of the
%supremum and infimum processes. In \sect{put} and \sect{call}, we
%use these representations to solve the optimal exercise problem
%for put-like and call-like options, respectively. In
%\sect{capexp}, timing a
%  capital expansion program is solved by reduction to a sequence
%  of investment opportunities of fixed size.

\section{L\'evy processes}\label{levy}

\subsection{}Recall that a L\'evy process is a
process with stationary independent increments (for general
definitions, see e.g. Sato (1999)). A L\'evy process may have a
Gaussian component and/or pure jump component. The latter is
characterized by the density of jumps, which is called the L\'evy
density. We denote it by $F(dx)$. Also, a L\'evy process can be
completely specified by its L\'evy exponent, $\Psi$, definable
from the moment-generating function $E\left[e^{z
X_t}\right]=e^{t\Psi(z)}$ (we confine ourselves to the
one-dimensional case). If $X_t$ is a L\'evy process with finite
variation jump component, then the L\'evy exponent is given by
\begin{equation}\label{psi1}
\Psi(z)=b z+\frac{\sg^2}{2} z^2+\INT(e^{zy}-1)F(dy),
\end{equation}
where $\sg^2$ and $b$ are the variance and drift coefficient of
the Gaussian component, and $F(dy)$ satisfies
\[ \int_{\R\setminus \{0\}} \min\{1, |y|\}F(dy)<+\infty. \]
 Equation \eq{psi1} is a special
case of the L\'evy-Khintchine formula; for the general case, see
e.g. Sato (1999). In this paper, we will illustrate our general
results for the case of the L\'evy density
\begin{equation}\label{densjump}
F(dx)=\cp\lp e^{-\lp x}{\bf 1}_{(0, +\infty)}(x) dx + \cm(-\lm)
e^{-\lm x}{\bf 1}_{(-\infty, 0)}(x) dx,
\end{equation}
 where $\lp>0>\lm$, and
$c_\pm> 0$. Then
\begin{equation}\label{jumpdif}
\Psi(z)= \frac{\sg^2}{2}z^2+bz+\frac{\cp z}{\lp-z} +\frac{\cm z
}{\lm-z},
\end{equation}
where $\sg^2> 0$ and $b\in \R$ are the variance and drift of the
Gaussian component. The $\Psi(z)$ is analytic in the strip $\Re z
\in (\lm, \lp)$.

\subsection{} The  L\'evy exponent
appears when we calculate the action of the infinitesimal
generator of $X_t$, denoted $L$, on exponential functions:
$Le^{zx}=\Psi(z)e^{zx}$. The EPV-operator \eq{epvX} calculates the
expected present value of the dividend stream $\de$. From the
fundamental relation between the infinitesimal generator and the
resolvent,
\begin{equation}\label{infgenres}
(q-L)\UqX=\UqX (q-L)=I,
\end{equation}
one concludes that $\UqX$
 acts on exponential functions as the
multiplication operator by the number $(q-\Psi(z))^{-1}$:
\begin{equation}\label{actUqX}
U^q_X e^{zx}=(q-\Psi(z))^{-1}e^{xz}.
\end{equation}
Let  $\barX_t=\sup_{0\le s \le t}X_s $ and $\uX_t=\inf_{0\le s\le
t}X_s$ be the supremum and infimum processes of $X_t$. Introduce
EPV-operators $\UqbX$ and $\UquX$ of the supremum and infimum
processes by
\[
\UqbX g(x):=E^x\left[\int_0^\infty e^{-qt} g(\barX_t)dt\right]:=
E\left[\int_0^\infty e^{-qt} g(\barX_t)dt\ |\ X_0=x\right]
\]
and
\[
\UquX g(x):=E^x\left[\int_0^\infty e^{-qt} g(\uX_t)dt\right] :=E\left[\int_0^\infty e^{-qt} g(\uX_t)dt\ |\
X_0=x\right],
\]
respectively.
% The $\UqbX$ and $\UquX$ are defined by the same
%formula as
% the
%resolvent of a Markov process, and we believe that the names {\em
%resolvents} of the supremum and infimum processes, respectively,
%are quite appropriate. In the economic applications, the name {\em
%expected present value operators} (which calculate the EPV of a
%stream $g$ under $X$, $\barX$ or $\uX$) seems to be quite
%appropriate, too.
It is straightforward to check that $q\UqbX$ and $q\UquX$ also act
on an exponential function $e^{zx}$ as multiplication operators by
certain numbers, which we denote $\kapq(z)$ and $\kamq(z)$,
respectively: \begin{equation}\label{actpm} q\UqbX
e^{zx}=\kapq(z)e^{zx},\quad  q\UquX e^{zx}=\kamq(z)e^{zx}.
\end{equation}
These numbers are
\begin{eqnarray}\label{actUqbX}
\kapq(z)&=&qE\left[\int_0^\infty e^{-qt} e^{z\barX_t}dt\right],\\
\label{actUquX}
 \kamq(z)&=&qE\left[\int_0^\infty e^{-qt}e^{z\uX_t}dt\right].
\end{eqnarray}
Notice that $\kapq(z)$ (resp., $\kamq(z)$) is analytic on the
half-plane $\Re z<0$ (resp., $\Re z>0$), and continuous up to the
boundary.
 The Wiener-Hopf factorization formula reads
 (see, e.g., Sato (1999), Section 45)
\begin{equation}\label{wh0}
\frac{q}{q-\Psi(z)}=\kapq(z)\kamq(z).
\end{equation}
 By applying $\UqX$, $\UqbX$ and $\UquX$ to an
exponential function $g(x)=e^{zx}$ and using \eq{actUqX} and
\eq{actpm}--\eq{wh0}, we obtain
\begin{equation}\label{wh}
\UqX g(x)=q\UqbX \UquX g(x)= q\UquX \UqbX g(x).
\end{equation}
By linearity, \eq{wh} holds for linear combinations of exponents,
and integrals of exponents, hence, for wide classes of functions.

Equation \eq{wh} means that  the EPV-operator of a L\'evy process
admits a factorization into a product of the EPV-operator of the
supremum process and the one of the infimum process.

\subsection{} For diffusions with exponentially distributed jumps,
$q-\Psi(z)$ is the rational function, which has 4 real zeroes; two
of them are positive, and two negative. We will call them $\bem_j$
and $\bep_j$, $j=1,2$, respectively. It is easy to show that $\lm$
separates the negative roots, and $\lp$ -- the positive ones. We
have $\bemt<\lm<\bepo<0<\bepo<\lp<\bept$. Since $q-\Psi(z)$ is
rational, the factors $\ka^\pm(z)$ can easily be obtained by
representing the LHS in \eq{wh0} as the fraction of two
polynomials, factorizing these polynomials out, and collecting the
factors with positive (respectively, negative) zeroes. For details
of these calculations and calculations below, see Levendorski\v{i}
(2004). We obtain
\begin{eqnarray}\label{kapq}
\kapq(z)&=&\frac{\bepo\bept(\lp-z)}{(\bepo-z)(\bept-z)\lp}\\
&=&\sum_{j=1,2}\frac{\ap_j}{\bep_j-z},
\end{eqnarray}
where
 \begin{equation}\label{apf}
  \apo=\frac{\bepo\bept(\lp-\bepo)}{(\bept-\bepo)\lp}, \quad
  \apt=\frac{\bepo\bept(\lp-\bept)}{(\bepo-\bept)\lp},
  \end{equation}
 are positive, and
  \begin{eqnarray}\label{kamq}
\kamq(z)&=&\frac{\bemo\bemt(\lm-z)}{(\bemo-z)(\bemt-z)\lm}\\
&=&\sum_{j=1,2}\frac{\am_j}{\bem_j-z},
\end{eqnarray}
where
 \begin{equation}\label{amf}
  \amo=\frac{\bemo\bemt(\lm-\bemo)}{(\bemt-\bemo)\lm}, \quad
  \amt=\frac{\bemo\bemt(\lm-\bemt)}{(\bemo-\bemt)\lm}
  \end{equation}
are negative. The operators $q\UqbX$ and $q\UquX$ act as follows:
\begin{eqnarray}\label{actuqp}
q\UqbX u(x)&=&\sum_{j=1,2}\ap_j\int_0^\infty e^{-\bep_j
y}u(x+y)dy,\\\label{actuqm} q\UquX
u(x)&=&\sum_{j=1,2}(-\am_j)\int^0_{-\infty} e^{-\bep_j y}u(x+y)dy.
\end{eqnarray}
To see this, it suffices to insert $u(x)=e^{zx}$, and use the
definition of the numbers $\kapq(z), \kamq(z)$.

\subsection{} In applications, one needs to calculate the EPV's of
exponentially growing payoffs. To ensure  that such EPV's are finite, $\Psi$ must be defined not only on the
imaginary axis $i\R$ but on a strip of the form $\lm\le \Re z\le \lp$, where $\lm\le 0<\lp$. An equivalent
condition is $\Psi(z)<+\infty$, for all $z\in [\lm, \lp]$. Since $\Psi(0)=0$ and $q>0$, there exist $\lm\le
\sgm\le 0<\sgp\le \lp$ such that
\begin{equation}\label{fin}
q-\Psi(z)>0, \quad \sgm\le z\le \sgp.
\end{equation}
In this case, $q-\Psi(z)$ does not vanish in the strip $\Re z\in
[\sgm, \sgp]$, the domains of the definition of $\kapq$ and
$\kamq$ contain this strip, and the equality \eq{wh0} holds for
$z$ in the strip $\Re z\in [\sgm, \sgp]$.

\section{Main results}\label{main}

\subsection{} Assume that the dividend stream is a non-negative
non-decreasing piece-wise continuous function of the stochastic
factor $X_t$, which admits a bound
\begin{equation}\label{finde}
\de(x)\le C(1+e^{\sgp x}).
\end{equation}

{\em Example 3.1.} The asset pays no dividends when the underlying
stochastic factor (say, value of the firm) is too low: $\de(x)=0,\
x\le x_0$, and the dividends increase exponentially after the
factor crosses a certain level:
\begin{equation}\label{de1}
\de(x)=(e^x-e^{x_0})^+:=\max\{0, e^x-e^{x_0}\}.
\end{equation}

{\em Example 3.2.} The asset pays no dividends when the underlying
stochastic factor (say, value of the firm) is too low: $\de(x)=0,\
x\le x_0$, and the dividends increase not so fast after the factor
crosses a certain level:
\begin{equation}\label{de2}
\de(x)=(x-x_0)^+.
\end{equation}

{\em Example 3.3.} The asset pays no dividends when the underlying
stochastic factor (say, value of the firm) is too low: $\de(x)=0,\
x\le x_0$. When the critical level is crossed, the dividends
increase but eventually the growth slows and essentially stops:
\begin{equation}\label{de3}
\de(x)=(1-e^{-x})^+.
\end{equation}

{\em Example 3.4.} The asset pays dividends at a fixed rate when
the underlying process is within a certain range; when the process
arrives in the next range, the dividend rate changes by a jump:
\begin{equation}\label{de4}
\de(x)=\sum_{j} \de_j {\bf 1}_{[d_j, +\infty)}(x).
\end{equation}
The sum can be finite (in this case, the dividends are capped, as
in Example 3.3) or infinite, which allows for unbounded growth of
dividends.

{\em Example 3.5.} The dividends are paid in the constant
proportion to the firm's value: $ d(X_t)=d X_t $,
 but the value itself is an
increasing function of a L\'evy process, $Y_t$: $X_t=f(Y_t).$ If
$f(y)$ is convex for $y<y_0$, and concave for $y>y_0$, the process
$X_t$ may exhibit a mean-reverting feature.

{\em Example 3.6.} We can easily generalize Examples 3.1--3.4 by
using $\de(f(Y_t))$, where $f$ is an increasing function, and
$Y_t$ is a L\'evy process.

 In the following sections, we prove the following results  for wide
classes of L\'evy processes, including diffusions with
exponentially distributed jumps.

\subsection{} First, we consider the perpetual American call. Let $h^*$ be
the solution to the equation
\begin{equation}\label{thrc0}
w(x):=(\UquX \de)(x)-K=0.
\end{equation}
(The solution exists if $\de(x)$ is sufficiently large for large
$x$, and it is unique since $\de$ is monotone). Then $h^*$ is the
optimal exercise level for the perpetual American call on the
asset with the dividend stream $\de$. After $h^*$ is found, we
calculate the rational call price, for $x\le h_*$:
\begin{equation}\label{callprice1}
V^*(x)=(q\UqbX {\bf 1}_{[h^*,+\infty)}w)(x).
\end{equation}
For exponential jump-diffusions, we use \eq{actuqm} and rewrite
\eq{thrc0} as
\begin{equation}\label{thrc11}
q^{-1}\sum_{j=1,2}(-\am_j)\int^0_{-\infty} e^{-\bem_j
y}\de(x+y)dy-K=0.
\end{equation}
The solution to \eq{thrc11} can easily be found by standard
numerical methods. In particular, in Examples 3.1--3.4 (and in
many others), the integral in \eq{thrp11} can be calculated
explicitly, and we have to solve an algebraic equation. Consider,
for instance,  Example 3.1. Without loss of generality, set
$x_0=0$. Then for $x\le 0$, the LHS is $-K$, therefore the root is
on the positive half-axis. For $x>0$, we change the variables
\begin{equation}\label{thrc112}
q^{-1}\sum_{j=1,2}(-\am_j)e^{\bem_j x}\int^x_{-\infty} e^{-\bem_j
y}\de(y)dy-K=0,
\end{equation}
and calculate
\[
\int_{-\infty}^x e^{-\bem_j y}\de(y)dy=\int_0^x e^{-\bem_j
y}(e^y-1)dy=\frac{e^{(1-\bem_j)x}-1}{1-\bem_j}-\frac{e^{-\bem_jx}-1}{-\bem_j}.
\]
We see that the first term on the RHS of \eq{thrc112} is
\begin{eqnarray*}
&&q^{-1}\sum_{j=1,2}(-\am_j)\left\{\frac{e^x-e^{-\bem_jx}}{1-\bem_j}-\frac{1-e^{-\bem_jx}}{-\bem_j}\right\}\\
&=&q^{-1}\sum_{j=1,2}\left\{\frac{-\am_j}{1-\bem_j}e^x-\frac{\am_j}{\bem_j}
+\frac{\am_je^{-\bem_jx}}{(1-\bem_j)\bem_j}\right\}\\
&=&q^{-1}\left\{\kamq(1)e^x-1+\sum_{j=1,2}\frac{\am_j}{(1-\bem_j)\bem_j}e^{-\bem_jx}\right\}.
\end{eqnarray*}
Thus, the equation for $h^*$ is
\[
\kamq(1)e^x-1+\sum_{j=1,2}\frac{\am_j}{(1-\bem_j)\bem_j}e^{-\bem_j
x}=qK+1,\quad x>0,
\]
and it can be easily solved. The equation being solved, we
represent $w$ in \eq{thrc0} in the form
\[
w(x)=q^{-1}\sum_{j=1,2}(-\am_j) \int_{-\infty}^0 e^{-\bem_j
y}(\de(x+y)-\de(h_*+y))dy,
\]
and calculate the rational call price using \eq{actuqp}:
\begin{eqnarray*}
V^*(x)&=&\sum_{k=1,2}\ap_k\int_0^{+\infty}e^{-\bep_k y}{\bf
1}_{[h^*, +\infty)}(x+y)w(x+y)dy\\
&=&\sum_{k=1,2}\ap_k\int_{h^*-x}^{+\infty}e^{-\bep_k y}w(x+y)dy\\
&=&q^{-1}\sum_{j,k=1,2}\ap_k(-\am_j)\\
&&\cdot\int_{h^*-x}^{+\infty}\int^0_{-\infty}e^{-\bep_k
y-\bem_j z}[\de(x+y+z)-\de(h^*+z)]dzdy\\
&=&q^{-1}\sum_{j,k=1,2}\ap_k(-\am_j)e^{\bep_k(x-h^*)}\\
&&\cdot \int_0^{+\infty}\int^{0}_{-\infty}e^{-\bem_j
z-\bep_k y}[\de(h^*+y+z)-\de(h^*+z)]dzdy\\
&=q^{-1}&\sum_{k=1,2}\ap_k\cD^+_ke^{\bep_k(x-h^*)},
\end{eqnarray*}
where the constants
\[
\cD^+_k=\sum_{j=1,2}(-\am_j)\int_{0}^{+\infty}\int_0^{+\infty}e^{\bem_j
z-\bep_k y}[\de(h^*-z)-\de(h^*+y-z)]dzdy
\]
can be calculated quite easily in all Examples 3.1-3.4 (and in
many other examples).

\subsection{} Now we consider the put. Let $h_*$ be the solution to the equation
\begin{equation}\label{thrp0}
w(x):=K-(\UqbX \de)(x)=0.
\end{equation}
(The solution exists, if $\de(x)=0$ or  sufficiently small for $x$
in a neighborhood of $-\infty$, and it is unique, since $\de$ is
non-decreasing). Then $h_*$ is the optimal exercise level for the
perpetual American put on the asset with the dividend stream
$\de$. After $h_*$ is found, we calculate the rational put price,
for $x\ge h_*$:
\begin{equation}\label{putprice1}
V_*(x)=(q\UquX {\bf 1}_{(-\infty, h_*]}w)(x).
\end{equation}
For exponential jump-diffusions, we use \eq{actuqp} and rewrite
\eq{thrp0} as
\begin{equation}\label{thrp11}
K-q^{-1}\sum_{k=1,2}\ap_k\int_0^{+\infty} e^{-\bep_k
y}\de(x+y)dy=0.
\end{equation}
The solution to \eq{thrp11} can easily be found by standard
numerical methods. In particular, in Examples 3.1--3.4 (and in
many others), the integral in \eq{thrp11} can be calculated
explicitly, and we have to solve an algebraic equation. The
equation being solved, we represent $w$ in \eq{thrp0} in the form
\[
w(x)=q^{-1}\sum_{k=1,2}\ap_k \int_0^{+\infty}e^{-\bep_k
y}(\de(h_*+y)-\de(x+y))dy,
\]
and calculate the rational put price using \eq{actuqm}:
\begin{eqnarray*}
V_*(x)&=&\sum_{j=1,2}(-\am_j)\int^0_{-\infty}e^{-\bem_j y}{\bf
1}_{(-\infty, h_*]}(x+y)w(x+y)dy\\
&=&\sum_{j=1,2}(-\am_j)\int^{h_*-x}_{-\infty}e^{-\bem_j y}w(x+y)dy\\
&=&q^{-1}\sum_{j,k=1,2}\ap_k(-\am_j)\\
&&\cdot\int^{h_*-x}_{-\infty}\int_0^{+\infty}e^{-\bem_j
y-\bep_k z}[\de(h_*+z)-\de(x+y+z)]dzdy\\
&=&q^{-1}\sum_{j,k=1,2}\ap_k(-\am_j)e^{\bem_j(x-h_*)}\\
&&\cdot \int^{0}_{-\infty}\int_0^{+\infty}e^{-\bem_j
y-\bep_k z}[\de(h_*+z)-\de(h_*+y+z)]dzdy\\
&=q^{-1}&\sum_{j=1,2}(-\am_j)\cD^-_je^{\bem_j(x-h_*)},
\end{eqnarray*}
where the constants
\[
\cD^-_j=\sum_{k=1,2}\ap_k\int_{0}^{+\infty}\int_0^{+\infty}e^{\bem_j
y-\bep_k z}[\de(h_*+z)-\de(h_*-y+z)]dzdy
\]
can be calculated quite easily in  Examples 3.1-3.4 (and in many
other examples).

%\section{Exchange options}
%Consider a perpetual option to exchange the stream of dividends
%$\de_{1,t}$ for another stream of dividends, $\de_{2,t}$. We
%assume that that the first stream is an exponential of a L\'evy
%process, $X_1(t)$, and the other one admits a representation in
%the form $\de_2(X_1, X_t)=e^{X_1(t)}\de(X_2(t))$, where
%$\de(X_2(t))$ is a non-decreasing function of another L\'evy
%process, $X_2(t)$. To allow for correlation between the factors
%$X_1(t)$ and $X_2(t)$, we assume that under an EMM chosen by the
%market, $X(t)=(X_1(t), X_2(t))$ is a bivariate L\'evy process with
%the L\'evy exponent $\Psi(z)=\Psi(z_1, z_2)$.

\section{The Wiener-Hopf method and some
applications}\label{WH}

\subsection{} From now on, our standing assumption is that the L\'evy process
$X$ satisfies the (ACP)-property (for the definition, see, e.g.,
Bertoin (1996) and Sato (1999)). Fix $h\in \R$, and set
$\tau=\tau^-_h=\inf\{\ t\ |\ X_t<h\}$. For $z$ in the upper
right-plane $\Re z\ge 0$,
 consider functions
\[
f(z; x)=f(q, h, z; x)=E^x[\exp(zX_\tau-q\tau)],
\]
and
\[
f^1(z; x)=f^1(q, h, z; x)=\kamq(z)^{-1}q\UquX u(z; x),
\]
where $u(z; x)=u(h, z; x)=\bfo_{(-\infty, h]}(x)e^{zx}$.
\begin{lem}\label{aux1} For $x>h$, $f(z; x)=f^1(z; x)$.
\end{lem}
\begin{proof}
For a fixed $x$, both functions are analytic in the half-plane
$\Re z>0$, and continuous up to the boundary. Hence, it suffices
to prove the equality for $z\in [0, \eps]$, where $\eps$ is some
positive number. If $z=0$, then $\kamq(z)^{-1}=1$, and the
equality
\[
E^x[\exp(zX_\tau-q\tau)]=qE^x\left[\int_0^{+\infty}
e^{-qt}\bfo_{(-\infty, h]}(\uX_t)dt\right]
\]
holds. Thus, the lemma is proved in the case $z=0$. The proof for
small positive $z$ consists of the following steps: $f(z;\cdot)$
is RCLL on $(h; +\infty)$ (right continuous with left limits);
$f^1(z;\cdot)$ is RCLL on $(h; +\infty)$; the Laplace transforms
of these two functions are equal.

Function $f(0;\cdot)$ is $q$-excessive (Proposition 41.5 (ii) and
(viii) in Sato (1999)). Since $X$ satisfies the (ACP)-property, a
$q$-excessive function is lower semi-continuous (Theorem 41.5 (4)
in Sato (1999)), but $f(0;\cdot)$ is evidently non-increasing;
hence, $f(0;\cdot)$ and $f^1(0; \cdot)$ are RCLL on $(h,
+\infty)$. Consider sufficiently small $z>0$  so that
$q-\Psi(z)>0$. Introduce $\Psi_z(w):=\Psi(w+z)-\Psi(z)$. This is
the L\'evy exponent of $X_t$ under the Esscher transform of the
measure $\bQ$; denote the Esscher transform by $\bQ_z$. Let
$E^{\bQ}$ and $E^{\bQ_z}$ be the expectation operators under $\bQ$
and $\bQ_z$, respectively. We have
\[
f(z; x):=E^{\bQ;
x}[\exp(-q\tau^-_h+zX_{\tau^-_h})]=e^{zx}E^{\bQ_z;
0}[\exp(-(q-\Psi(z))\tau^-_h)].
\]
Since $X$ satisfies the (ACP)-property under $\bQ$, it satisfies
the (ACP) property under $\bQ_z$. Hence, the last factor on the
RHS is RCLL on $(h, +\infty)$, and  $f(z;\cdot)$, its product with
a continuous function, is RCLL on $(h, +\infty)$ as well.

 To prove that $f^1(z; \cdot)$ is
RCLL on $(h, +\infty)$, we change
 the variables $x\mapsto x+h$, so that $h$ becomes 0, and represent $f^1(z; \cdot)$
in the form
 \begin{equation}\label{f1f} f^1(z; x)=\kamq(z)^{-1}\left(q\UquX
\bfo_{(-\infty, 0]}(x)+q\UquX \bfo_{(-\infty,
0]}(e^{z\cdot}-1)(x)\right),\end{equation}
 The function $\bfo_{(-\infty,
0]}(x)(e^{zx}-1)$ is continuous, therefore the second term in the
brackets on the RHS in \eq{f1f} is continuous. The first term in
the brackets equals $f(0; x)$, hence it is RCLL on $(h, +\infty)$.
We conclude that the sum is RCLL  on $(h, +\infty)$.

%Since both functions are continuous on $(h, +\infty)$, we can
%prove the equality by considering the Laplace transforms.
Now we consider the Laplace transforms. The fluctuation identity
(3.13) in Hilberink--Rogers (2002) states that for any $\mu>z$,
\[
\int_0^\infty  e^{-\mu x} f(z;
x)dx=\frac{1}{\mu-z}\left[1-\frac{\kamq(\mu)}{\kamq(z)}\right],
\]
and  therefore we need to prove  the same equality for $f^1$. We
have
\[
f^1(z; x)=\kamq(z)^{-1}q\UquX\left(1-\bfo_{(0,
+\infty)}(x)\right)e^{zx}=e^{zx}-\kamq(z)^{-1}q\UquX v(z; x),
\]
where $v(z; x)=\bfo_{(0, +\infty)}(x)e^{zx}$. For $w$ in the upper
half-plane $\Re w>\Re z$, the Laplace transform
\[
\hat v(z; w)=\int_0^{+\infty} e^{-wx} v(z;
x)dx=\int_0^{+\infty}e^{-x(w-z)}dx=(w-z)^{-1}
\]
is well-defined; therefore, using the inverse Laplace transform,
we obtain, for any $\sg>\Re z$,
\begin{equation*}\label{f1}
f^1(z; x)=e^{zx}-(2\pi i
)^{-1}\int_{-i\infty+\sg}^{+i\infty+\sg}e^{wx}\kamq(z)^{-1}\kamq(w)(w-z)^{-1}dw.
\end{equation*}
We take $\mu>z$ (and $\sg\in (z, \mu)$), and calculate the Laplace
transform
\[ \int_0^{+\infty} e^{-\mu
x}f^1(z; x)dx=\frac{1}{\mu-z}-\frac{\kamq(\mu)}{\kamq(z)(\mu-z)}
=\frac{1}{\mu-z}\left[1-\frac{\kamq(\mu)}{\kamq(z)}\right].
 \]
  Thus, the Laplace
 transforms of $f(z; \cdot)$ and $f^1(z; \cdot)$ are equal, which
 completes the proof.
%\end{equation}
%The multiplication of \eq{f2} by $\mu$ finishes the proof.
\end{proof}
Below, we consider piece-wise continuous streams $g$; this
condition can be relaxed.
\begin{thm}\label{epvhp} Assume that the infimum process is non-trivial,
and $g$ is  non-negative. Then
\begin{equation}\label{whm}
E^x\left[e^{-q\tau^-_h}\UqX g(X_{\tau^-_h})\right]=q\UquX
\bfo_{(-\infty, h]}\UqbX g(x),\quad \forall\ x>h.
\end{equation}
\end{thm}
\begin{proof}
Consider first $g(x)=e^{zx}$, where $z\in i\R$.  Using \eq{actUqX}
and \eq{actpm}, we can rewrite \eq{whm} as
\[ (q-\Psi(z))^{-1}f(z; x)=q^{-1}\kapq(z)f^1(z; x),
\]
where $f$ and $f^1$ are the functions in \lemm{aux1}. This
equality holds on the strength of \eq{wh} and \lemm{aux1}, hence
\eq{whm} is proved for oscillating exponents.

 Next, consider $g\in
C_0^\infty(\R)$. We  represent $g$ as the Fourier integral, use
\lemm{aux1} under the integral sign, and obtain \eq{whm}. Finally,
a general piece-wise continuous $g$ can be approximated by a
sequence $\{g_n\}\in C_0^\infty(\R)$, which converges to $g$
point-wise, from below. For each $n$, \eq{whm} with $g_n$ instead
of $g$ holds. By the Monotone Convergence Theorem,  the LHS and
RHS with $g_n$ instead of $g$ have point-wise limits, which are
the LHS and RHS with $g$.
\end{proof}

\subsection{} Fix $h\in \R$, and set $\tau=\tau^+_h=\inf\{\ t\ |\
X_t>h\}$. If we change the direction of the real line, a
neighborhood of $-\infty$ becomes a neighborhood of $+\infty$, the
supremum process becomes the infimum process, and vice versa.
Hence, by symmetry, we obtain
\begin{thm}\label{epvhm} Assume that the supremum process is non-trivial,
and $g$ is non-negative. Then
\begin{equation}\label{whp}
E^x\left[e^{-q\tau^+_h}\UqX g(X_{\tau^+_h})\right]=q\UqbX
\bfo_{[h, +\infty)}\UquX g(x),\quad \forall\ x <h.
\end{equation}
\end{thm}
\subsection{} Let \eq{fin} hold.
 If $g$ is piece-wise continuous and satisfies an estimate
 \begin{equation}\label{estm}
 |g(x)|\le Ce^{\sgm x}, \quad x<h,
 \end{equation}
 then the LHS in \eq{whm} is finite, and \eq{whm} holds.
 Similarly, if $g$ is piece-wise continuous and
\begin{equation}\label{estp}
 |g(x)|\le Ce^{\sgp x}, \quad x>h,
 \end{equation}
 then the LHS in \eq{whp} is finite, and \eq{whp}
  holds.

  \section{Optimal exercise boundary and rational price of a
  perpetual put-like option}\label{put}
  \subsection{} We use the following general lemma (Lemma 5.1 in
  Boyarchenko and Levendorski\v{i} (2002b) and Lemma 7.1 in Boyarchenko and Levendorski\v{i}
  (2002b)). We formulate it for a special case of a process
  on the line, and an optimal stopping region of the form
  $(-\infty, h]$.
  \begin{lem}\label{free}
  Let $h_*$ and a  function $V_*$ satisfy the following conditions:
  \begin{eqnarray}\label{freq1}
  (q-L)V_*(x)&=&0,\quad x>h_*;\\ \label{freq2}
V_*(x)&=&G(x), \quad x\le h_*;\\ \label{frineq1} V_*(x)&\ge
&\max\{G(x), 0\}, \quad x\in\R;\\ \label{frineq2} (q-L)V_*(x)&\ge
&0,\quad x<h_*;\\ \label{frreg1}
 W_*&:=& (q-L)V_*\quad {\rm is\ universally\ measurable}; \\ \label{frreg2}
 \UqX W_*&=& V_*.
 \end{eqnarray}
 Then the pair $(\tau^-_{h_*}$, $V_\ast)$ is the solution to the optimal
 stopping problem \eq{opt} in the class $\cM$.
 \end{lem}
 \subsection{}
 The optimal stopping problem for a put-like option is trivial if the infimum
 process is trivial. Hence, we presume in this section that the infimum
 process is non-trivial.

 Fix arbitrary $h$.  If $X$ satisfies the (ACP)--property, and $G$ is continuous and does not grow too
 fast at infinity, so that $ V^-_0(h; x)=E^x\left[e^{-q\tau^-_h}G(X_{\tau^-_h})\right]$ is finite
(a sufficient condition
 is \eq{estm}), then \\ $V^-_0(h; x)$ satisfies \eq{freq1} with $h$ instead of $h_*$
 (see Theorem 2.12 and Remark 2.1 in Boyarchenko and Levendorski\v{i} (2002b)
 or   Theorem 2.1 in Boyarchenko and Levendorski\v{i} (2002c)), and clearly,
 $V^-_0(h; x)$ satisfies \eq{freq2} for $x<h$. By \theo{epvhp}, for $x>h$,
 $V^-_0(h; x)=V^-(h; x):=q\UquX \bfo_{(-\infty, h]}\UqbX g(x)$. To prove the
  equality $V^-_0(h; x)=V^-(h; x)$ for $x<h$,
 we use \eq{wh} to represent $V^-(h; x)$ in
the form
\begin{eqnarray}\nonumber
V^-(h; x) &=&q\UquX \UqbX g(x)-q(\UquX {\bfo}_{(h,
+\infty)}w)(x)\\\nonumber &=& \UqX g(x) - v(h; x)\\\label{Vv} &=&
G(x) - v(h; x),
\end{eqnarray}
where $v(h; x)=q(\UquX {\bfo}_{(h, +\infty)}w)(x)$. For $x\le h$,
\[
v(h; x)=qE\left[\int_0^{+\infty} e^{-qt}({\bfo}_{(h,
+\infty)}w)(\uX_t)\ |\ X_0=x\right]=0,
\]
therefore $V^-(h; x)=G(x), x\le h$. But for $x< h$,
$E^x\left[e^{-q\tau^-_h}\UqX g(X_{\tau^-_h})\right]=G(x)$ as well.

It remains to find
 $h_*$ such that $V^-(h; x)$ is continuous, and satisfies \eq{frineq1}--\eq{frreg2}.
 Assume that $g$ is piece-wise continuous, and satisfies \eq{estm}. $X$ satisfies the (ACP)-property,
 hence $G=\UqX g$ is continuous, and $G$ satisfies \eq{estm}, since $g$ does. Introduce
 $w=\UqbX g$, and assume that $w$ is a continuous function that
 satisfies
 \begin{equation}\label{suff0m}
 w\ {\rm changes\ sign\ from\ ``+"\ to\ ``-",\ and\ only\ once.}
 \end{equation}
 If $g$ is continuous, $w$ is continuous as well, and a sufficient condition for \eq{suff0m} is that $g$ is
 decreasing.
 %\begin{equation}\label{suffm}
 %g\ {\rm is\ decreasing}.
 %\end{equation}
% Notice that \eq{suffm}
This condition makes a perfect economic sense for a perpetual
put-like option: the stream
 of payoffs increases when the stochastic factor decreases.

 Denote
 by $h_*$ the solution of the equation
 \begin{equation}\label{thrm1}
 w(x)=\UqbX g(x)=0,
\end{equation}
and set $V_*(\cdot)=V^-(h_*; \cdot)$.
\begin{thm}\label{optput}
Let \eq{suff0m} hold, and $g$ be non-increasing. Then
$(\tau^-_{h_*}, V_*)$ is the optimal solution to the stopping
problem, in the class $\cM$.
\end{thm}
\begin{proof}  Due to the choice of $h_*$, $\bfo_{(-\infty,
h_*]} w$ is continuous and non-negative. Therefore, $V^-(h^*;
\cdot)$ is continuous, and
\begin{eqnarray}\nonumber
V^-(h_*; x)&=&q\UquX \bfo_{(-\infty, h_*]} w(x)\\\label{opt0}
&=&qE\left[\int_{0}^{+\infty} e^{-qt} \left(\bfo_{(-\infty,
h_*]}w\right)(x+\uX_t)dt\right]\ge 0, \quad \forall\ x.
\end{eqnarray}
Further, consider \eq{Vv} with $h=h_*$, the solution to
\eq{thrm1}. Due to \eq{suff0m}, $\bfo_{(h_*, +\infty)}w$ is
non-positive, hence $v(h_\ast; \cdot)$ in \eq{Vv} is non-positive,
and we conclude from \eq{Vv} that $V^-(h_\ast; x)\ge G(x)$ for all
$x$.
 Thus, $V_*$ satisfies \eq{frineq1}. Since \eq{freq1} holds, we
need to check \eq{frineq2} on $(-\infty, h_*)$. Below, we will
show that
\begin{equation}\label{Wnoninc}
W_*=(q-L)V_*\quad {\rm is\ non-increasing\ on}\ (-\infty, h_*), \
\end{equation}
and
\begin{equation}\label{Wlim}
W_*(h_*-0)\ge 0.
\end{equation}
Conditions \eq{frineq2} and \eq{frreg1} follow immediately from
\eq{Wnoninc} and \eq{Wlim}, and it remains to check \eq{frreg2}.
Since $\bfo_{(-\infty, h_*]}\UqbX g$ is continuous,
$V_*=V^-(h_\ast; \cdot)$ given by the LHS in \eq{whm} is
continuous, and since $W_*$ is universally measurable and $X$
satisfies the (ACP)-property, $\UqX W_*$ is continuous. Therefore
it suffices to prove \eq{frreg2} in the sense of generalized
functions:
\[
\INT \UqX W_*(x)u(x)dx=\INT V_*(x) u(x)dx,\quad \forall\ u\in
C_0^\infty(\R).
\]
By the standard duality argument,
\[
\INT \UqX (q-L)V_*(x)u(x)dx=\INT V_*(x)(q-\tilde L)U^q_{\tilde
X}u(x)dx,
\]
where $\tilde X$ is the dual process and $\tilde L$ its generator.
Since $u\in C^\infty_0(\R)$, $(q-\tilde L)U^q_{\tilde X}u=u$, and
the proof of \eq{frreg2} is finished.

It remains to prove \eq{Wnoninc} and \eq{Wlim}. Represent $W_*$ in
the form
\[
W_*=(q-L)G-(q-L)w^+,
\]
where $w^+=q\UquX\bfo_{(h_*, +\infty)}w$. Since ${\rm supp}
w^+\subset [h_*, +\infty)$, and the Gaussian part of the
infinitesimal generator is a local (differential) operator, we
have for $x<h_*$
\begin{eqnarray*}
W_*(x)&=&(q-L)G(x)\\
&&+\INT\left(w^+(x+y)-w^+(x)-\bfo_{[-1,1]}(y)(w^+)'(x)\right)F(dy)\\
&=&g(x)+\int_{h_*-x}^{+\infty}w(x+y)F(dy),
\end{eqnarray*}
where $F(dy)$ is the L\'evy density. Since $(q-L)G=g$ is
non-increasing,  $w=\UqbX g$ is non-increasing as well, hence both
terms on the RHS are non-increasing, and \eq{Wnoninc} is proved.

Finally, assume that \eq{Wlim} fails. On the strength of
\eq{Wnoninc}, $W_*$ must be negative on some interval $(h, h_*)$,
where $h<h_*$. By applying $\UqbX$ to $W_*=(q-L)V_*$ and using
\eq{whm} and \eq{wh}, we obtain
 \[
 \UqbX W_*=\UqbX (q-L) q\UquX \bfo_{(-\infty, h_*]}w=\bfo_{(-\infty,
 h_*]}w.\]
For $x\in (h, h_*)$, we have $\UqbX W_*(x)\le 0$, but
$\bfo_{(-\infty, h_*]}w(x)>0$, a contradiction. Thus, \eq{Wlim}
holds, and the proof is complete.
\end{proof}
\subsection{} In the case of the put-option on a stock that pays
no dividends, we have $q=r$, and the straightforward application
of the above scheme faces evident difficulty since $K-e^x$ cannot
be represented as the EPV of any stream of  payoffs. Indeed, $e^x$
is an eigenfunction of $q-L$: $(q-L)e^x=0$. Since the put option
will not be exercised if $K-e^x$ is negative, we can try to
overcome this difficulty by choosing a sufficiently smooth $G$
which coincides with $K-e^x$ on $(-\infty, \log K]$, is negative
on $(\log K, +\infty)$, and does not grow (in absolute value) too
fast as $x\to+\infty$, so that  $g=(q-L)G$ satisfies $G=\UqX g$.
However, in this case we need to ensure that $w=\UqbX g=
(q\UquX)^{-1} G$ changes sign only once, and $(q-L)q\UquX
\bfo_{(-\infty, h_*]}w$ is monotone. Instead, we reformulate the
conditions of \theo{optput} in terms of the payoff function $G$
and $w=(q\UquX)^{-1}G$ so that we avoid the use of a stream $g$
altogether. Notice that $(q\UquX)^{-1}G$ is easily calculated if
$G$ is a linear combination of exponential functions since
$(q\UquX)^{-1}e^{zx}=\kamq(z)^{-1}e^{zx}$.
 In the
proof of \theo{optput}, the following conditions are essential
(and even these conditions can be relaxed):
\begin{equation}\label{suffm2}
(q-L)G\quad {\rm is\ non-increasing\ on}\ (-\infty, h_*),
\end{equation}
and
\begin{equation}\label{suffm3}
w=(q\UquX)^{-1}G\quad {\rm is\ non-increasing\ on}\ (h_*,
+\infty).
\end{equation}
Set $V_*(x)=q\UquX \bfo_{(-\infty, h_*]} w(x)$.
\begin{thm}\label{optput2}
Let $G$ be continuous, let $w=(q\UquX)^{-1}G$ satisfy \eq{suff0m},
and let \eq{suffm2} and \eq{suffm3} hold. Then the pair
$(\tau^-_{h_*}, V_*(\cdot))$  is the optimal solution to the
optimal stopping problem, in the class $\cM$.
\end{thm}
{\em Example.} For the perpetual American put without dividends,
$G(x)=K-e^x$, $w(x)=K-\kamq(1)^{-1}e^x$, and $(q-L)G(x)=qK$.
Clearly, the conditions of \theo{optput2} are satisfied, and the
optimal exercise log-price is $h_*=\ln(K\kamq(1))$.

\subsection{} In Boyarchenko and Levendorski\v{i} (2000,  2002a,b), we
 introduced a wide class of L\'evy processes: Regular L\'evy
Processes of Exponential type (RLPE), and, in particular, solved
the optimal stopping problem for perpetual American options on a
stock driven by an RLPE. A L\'evy process is an RLPE if its L\'evy
density exhibits a regular growth near 0 and exponential decay at
infinity. The class of RLPE includes several families of L\'evy
processes used in empirical
 studies of financial markets: jump-diffusion processes
with exponentially distributed jumps, hyperbolic processes, and
extended Koponen's family of Truncated L\'evy processes or KoBoL
processes (later used under the name CGMY-model in Carr et al.
(2002)). For details, see Boyarchenko and Levendorski\v{i}
(2002a,b,c). Due to certain analytical complications, variance
gamma processes were not treated in Boyarchenko and
Levendorski\v{i} (2000,  2002a,b), although L\'evy processes with
non-trivial gaussian and variance gamma components were allowed.
%\footnote{
% The class RLPE includes several families of L\'evy processes used in empirical
% studies of financial markets: jump-diffusion processes
%with exponentially distributed jumps, Hyperbolic processes
%constructed by Eberlein et al (1995, 1998), Normal Inverse
%Gaussian processes introduced by Barndorff-Nielsen (1998), and
%extended Koponen's family of Truncated L\'evy processes or KoBoL
%processes. Koponen (1995) considered the case of symmetric L\'evy
%densities, and the generalization for the case of non-symmetric
%densities was given in Boyarchenko and Levendorski\v{i} (2000,
%2001, 2002a, b) (in Boyarchenko and Levendorski\v{i} (2002a, b),
%the name KoBoL processes is used); we also derived a more
%convenient formula for the characteristic exponent. Later, Carr et
%al (2002) used the simplest version of formula (with the same
%formula for the characteristic exponent but different labels for
%the parameters) under the name CGMY-model.}
In Boyarchenko and Levendorski\v{i} (2000) we found the exercise
boundary for the perpetual American put and rational put price,
and in Boyarchenko and Levendorski\v{i} (2002a,b), we obtained
similar results for calls and put-like and call-like options with
more general payoff functions of the form $\max\{G(X_t), 0\}$. For
a general $G$, the optimality of the solution in the class $\cM_0$
was shown; the optimality in the class $\cM$ was proved for $G$ a
linear combination of exponential functions, satisfying certain
conditions. The results were formulated in terms of the factors in
the Wiener-Hopf factorization formula, or, equivalently, in terms
of the supremum and infimum processes, and we conjectured that
they held for any L\'evy process. Later, for the case of the put
and call option, Mordecki (2002) considered arbitrary L\'evy
process. He showed that the value functions in appropriate
discrete time models have the limit as the time step goes to 0,
and finished the proof by a sentence which stated that the
optimality of the limit ``\ldots can be proved exactly as in the
discrete time model in Darling et al (1972)"; the precise meaning
of the word ``exactly" is not clear in this context. It seems
unlikely that the proof in Darling et al (1972), which uses the
sufficient conditions in terms of the transition operator of the
random walk, can be repeated word by word in the case of a L\'evy
process, where the verification is typically more involved (see,
e.g., Boyarchenko and Levendorski\v{i} (2002a, b), and references
therein and in Mordecki (2002)). Thus, the proof in Mordecki
(2002) is incomplete. Essentially, it is a form of a reasonable
guess of the answer, for a particular case of the puts and calls;
but Boyarchenko and Levendorski\v{i} (2002a, b) obtained the
optimal solution in general terms earlier (albeit not for any
L\'evy process), for wider classes of payoffs.

\section{Optimal exercise boundary and rational price of a
  perpetual call-like option}\label{call}
  The statements, arguments, and proofs are mirror reflections of
  the ones for put-like options. We assume that the supremum
  process is non-trivial, and
 $g$ is piece-wise continuous, and satisfies \eq{estp}. Since $X$
satisfies the (ACP)-property,
 $G=\UqX g$ is continuous, and it satisfies \eq{estp}. Introduce
 $w=\UqbX g$, and assume that $w$ is a continuous function that
 satisfies
 \begin{equation}\label{suff0p}
 w\ {\rm changes\ sign\ from\ ``-"\ to\ ``+",\ and\ only\ once.}
 \end{equation}
 If $g$ is continuous, $w$ is continuous as well, and a sufficient condition for \eq{suff0p} is that $g$ is
 increasing.
 %\begin{equation}\label{suffp}
 %g\ {\rm is\ increasing}.
 %\end{equation}
 %Notice that \eq{suffp}
 This condition makes a perfect economic sense for a perpetual call-like option: the stream
 of payoffs increases when the stochastic factor increases.

 Denote
 by $h^*$ the solution to the equation
 \begin{equation}\label{thrp1}
 w(x)=\UquX g(x)=0,
\end{equation}
and set $V^*(x)=q\UqbX \bfo_{[h^*, +\infty)}w(x)$.
\begin{thm}\label{optcall}
Let $g$ be non-decreasing,  and \eq{suff0p} hold. Then the pair $(\tau^+_{h^*}, V^*(\cdot))$ is the optimal
solution to the stopping problem, in the class $\cM$.
\end{thm}
{\em Example.} Assume that $q-\Psi(1)>0$, and consider the perpetual call option on a dividend paying stock. We
have $\kapq(1)<+\infty$,
 $G(x)=e^x-K$, $g(x)=(q-L)G(x)=(q-L)e^x-qK$, $w(x)=(q\UqbX)^{-1}G(x)=\kapq(1)^{-1}e^x-K$.
Clearly, $g$ is increasing, condition \eq{suff0p} holds, and the optimal exercise log-price is
$h^*=\ln(K\kapq(1))$.

Similarly to \theo{optput2}, we can replace the condition that $g$ is non-decreasing with the following pair of
conditions:
\begin{equation}\label{suffp2}
(q-L)G\quad {\rm is\ non-decreasing\ on}\ (h^*, +\infty),
\end{equation}
and
\begin{equation}\label{suffp3}
(q\UquX)^{-1}G\quad {\rm is\ non-decreasing\ on}\ (-\infty, h^*).
\end{equation}
\begin{thm}\label{optcall2}
Let $G$ be continuous, let $w=(q\UquX)^{-1}G$ satisfy \eq{suff0p},
and let \eq{suffp2} and \eq{suffp3} hold. Then the pair
$(\tau^+_{h_*}, V^*(\cdot))$ is the optimal solution to the
stopping problem, in the class $\cM$.
\end{thm}

\section{Incremental capital expansion}\label{capexp}
Consider a firm whose production function depends only on capital
$G=G(K)$. We assume
 that $G(K)$
differentiable, concave, and satisfies the Inada conditions; the revenue flow is $P_tG(K_t)$, where $P_t$ is the
spot price of the firm's output. A similar situation was considered in Dixit and Pindyck (1996) for the geometric
Brownian motion model, and extended by Boyarchenko (2004) for geometric L\'evy processes. In those papers, the
price of the firm's output was modeled as $P_t=e^{X_t}$, where $X_t$ is the Brownian motion and a L\'evy process,
respectively. In the present paper, we consider more general case, when $P_t=P(X_t)$ is an increasing function of
a stochastic factor $X_t$, which follows a L\'evy process. In particular, such a payoff may account for the case
when the firm chooses both capital and costlessly adjustable labor as in Abel and Eberly (1999) for the gaussian
model, or in Boyarchenko and Levendorski\v{i} (2004) for the discrete time model. Should the firm decide to invest
a unit of capital, it suffers the installation cost $C$. At the end of this Section, we allow for a stochastic
operational cost as well. In order not to change the notation of the previous Sections, we denote the riskless
rate by $q$. The firm's objective is to chose the optimal investment strategy $\cK=\{K_t, t\ge 0\}, K_0=K, X_0=x$,
which maximizes the NPV of the firm:
\begin{equation}\label{vf1}
V(K, x)= \sup_{\cK}E^x\left[\int_{0}^{+\infty}
e^{-qt}(P(X_t)G(K_t)-
 qC K_t)dt\right].
 \end{equation}
%Here we treat the current level $x$ of the stochastic factor and
%capital stock $K$ as state variables, and $\cK$ as a  control
%variable. Due to irreversibility of investment, $K_t$ is
%non-decreasing in $t$.

To ensure that firm's value \eq{vf1} is bounded, we impose a resource constraint: there exists $\bar{K}<\infty$,
such that $K_t\le\bar{K}, \ \forall t$. Also we assume that $X$ satisfies \eq{fin}, and the function $P$ satisfies
\eq{estp}. For the case $P(X_t)=e^{X_t}$, the last two conditions reduce to $q-\Psi(1)>0$. These conditions and
properties of the production function ensure that the value function \eq{vf1} is well defined.

Formally, the manager has to choose both the timing and the size
of the capital expansion. However, it is well-known (see, for
example, Dixit and Pindyck (1996)) that for each level of the
capital stock, it is only necessary to decide when to invest. The
manager's problem is equivalent to finding the boundary (the
investment threshold), $h(K;C)$, between two regions in the state
variable space $(K,x)$: inaction and action ones. For all pairs
$(K, x)$ belonging to the inaction region, it is optimal to keep
the capital stock unchanged. In the action region, investment
becomes optimal. To derive the equation for the investment
boundary, suppose first that every new investment can be made in
chunks of capital, $\Delta K$, only\footnote{The authors are
indebted for this simplifying trick to Mike Harrison; the initial
proof in Boyarchenko (2004) was more involved.}. In this case, the
firm has to suffer the cost $C\Delta K$, and the EPV of the
revenue gain due to this investment can be represented in the form
of the EPV of the stream $g(X_t)=(G(K+\Delta K)-
G(K))P(X_t)-qC\Delta K$. On the strength of the result of Section
4, the optimal exercise boundary is determined from the equation
$\UquX g(h)=0$. In the geometric L\'evy case, which we consider
first in order to simplify the presentation of the main idea of
the proof, the equation for the threshold  can be written as
\begin{equation}\label{marsh1}
q^{-1}(G(K+\Delta K)- G(K))\kamq(1)e^h=C\Delta K.
\end{equation}
 Dividing by $\Delta K$ in \eq{marsh1}
and passing to the limit, we obtain the equation for the optimal
threshold, $h^\ast=h^\ast(K)$:
\begin{equation}\label{marsh2}
\kamq(1)G'(K)e^h=qC. \end{equation} Equivalently, the optimal
exercise price is
\begin{equation}\label{marsh3}
e^{h^\ast}=e^{h^\ast(K)}=\frac{qC}{\kamq(1)G'(K)}.
\end{equation}
For the rigorous justification of this limiting argument, see Boyarchenko (2004). Let $h=h(K; \Delta)$ be the
solution to \eq{marsh1}. Then  the option value  associated with the chunk of capital $\Delta K$, at the price
level $e^x$, is
\[
\UqbX \bfo_{[h, +\infty)}(x) ((G(K+\Delta K)- G(K))\kamq(1)e^x-qC\Delta K).
\]
As $\Delta K\to 0$, we have $h=h(K; \Delta)\to h^\ast(K)$;
therefore, dividing by $\Delta K$ and passing to the limit, we
obtain the formula for the derivative of the option value of
future investment opportunities w.r.t. $K$:
\begin{equation}\label{valmarsh1}
V^{\mathrm{opt}}_K(K, x)=\UqbX \bfo_{[h^\ast, +\infty)}(x)
(G'(K)\kamq(1)e^x-qC).
\end{equation}
Substituting $C$ from \eq{marsh3} into \eq{valmarsh1} and using the definition of $\kamq(1)$, we obtain
\begin{eqnarray*} V^{\mathrm{opt}}_K(K, x)&=&qE\left[\int_0^\infty e^{-qt}G'(K)e^{\uX_t}dt\,\vert\,
X_0=0\right]\\ && \times E^x\left[\int_0^\infty e^{-qt}(e^{\barX_t}-e^{h^*})_+dt\right].\end{eqnarray*} The last
formula factors out the contributions of the infimum and supremum price processes to the marginal option value of
capital. The first expectation on the RHS decreases if the probability of downward jumps in prices increases, and
the second expectation increases if the probability of positive jumps in prices increases. Hence the marginal
option value of capital increases in downward uncertainty and decreases in upward uncertainty. The overall effect
of uncertainty is ambiguous.

Notice that the proof of \eq{marsh2} in Boyarchenko (2004) was based on the reduction to the case of the perpetual
American call, and therefore the generalization for more general dependence on the stochastic factor was not
possible. Here the result holds for any continuous increasing revenue flow $R(K, x)$, and  the formula for the
optimal investment threshold obtains in the form:
\begin{equation}\label{marsh4}
\UquX  R_K(K, h)=C,
\end{equation}
where the EPV--operator $\UquX$  acts w.r.t. the second argument.
For instance, if the firm faces the operational cost $a+bK
e^{X_t/2}$, then the revenue flow is $R(K_t,
X_t)=e^{X_t}G(K_t)-a-bKe^{X_t/2}$, and instead of \eq{marsh3}, we
now have
\begin{equation}\label{marsh5}
q^{-1}[\kamq(1)G'(K)e^{h}-b\kamq(1/2)e^{h/2}]-C=0.
\end{equation}
The function on the LHS in \eq{marsh5} changes sign only once, and
therefore the solution to equation \eq{marsh5} gives the optimal
investment threshold.

One can also consider non-exponential dependence of the price on
the stochastic factor.

Equation \eq{marsh4} says that it is optimal to increase the capital stock the first time the EPV of the marginal
revenue, calculated under the assumption that the underlying stochastic process is replaced by the infimum
process, reaches or overshoots the marginal cost of investment. This rule reflects and extends the bad news
principle spelled out by Bernanke (1983).

\end{document}